\documentclass[conference]{IEEEtran}
\ifCLASSINFOpdf
\else
\fi
\usepackage[english]{babel}
\usepackage[T1]{fontenc}

\usepackage[utf8x]{inputenc}
\usepackage{cite}
\usepackage{graphicx}
\usepackage{color}
\usepackage{amssymb}
\usepackage{amsthm}
\usepackage{amsxtra}
\usepackage{amsmath}
\usepackage{multirow}
\usepackage{epsfig}
\usepackage{comment}
\usepackage{psfrag}
\usepackage{url}
\usepackage{textcomp}
\usepackage{enumerate}
\usepackage[printonlyused]{acronym}
\usepackage{balance}
\usepackage{algorithm}
\usepackage{algpseudocode}
\usepackage{nomencl}
\usepackage{units}
\usepackage{subcaption}
\usepackage{graphicx}
\usepackage{pstricks, pst-plot, pst-grad, pst-node, pstricks-add}
\usepackage{pifont}
\usepackage{glossaries}
\usepackage{siunitx}

\newrgbcolor{darkgreen}{0.2 0.6 0.2}
\newrgbcolor{darkred}{0.6 0.2 0.2}
\newrgbcolor{darkblue}{0.2 0.2 0.6}
\newrgbcolor{purple}{0.2 0.1 0.9}
\begin{document}

\title{Error Convergence Analysis and Stability of a Cloud Control AGV}



%

\author{
\IEEEauthorblockN{Shreya Tayade\IEEEauthorrefmark{1},
Peter Rost\IEEEauthorrefmark{2} and
Andreas Maeder\IEEEauthorrefmark{2}}
\IEEEauthorblockA{\IEEEauthorrefmark{1}Intelligent Networks Research Group, \\ German Research Center for Artificial Intelligence, Kaiserslautern, Germany\\
Email: Shreya.Tayade@dfki.de}
\IEEEauthorblockA{\IEEEauthorrefmark{2}Nokia Bell Labs,
Munich, Germany\\
Email: \{peter.m.rost, andreas.maeder\}@nokia-bell-labs.com}
}
\maketitle

\begin{abstract}
In this paper, we present a cloud based Automated Guided vehicle (AGV) control system. A controller in an Edge cloud sends the control inputs to an AGV to follow a predefined reference track over a wireless channel. The AGV feedback the position update via uplink channel. The objective of this paper is to evaluate the stability criterion of an AGV control system in presence of an uplink channel outages. Moreover, we also analyse the impact of feedback control parameters on the error convergence. The results show error convergence at higher rate with optimal selection of feedback parameters. The optimal feedback parameters that converges the error with critical damping is evaluated for two scenarios; with limited AGV velocity and without the limitation on AGV velocity. Furthermore, the paper describe the discretization process of a continuous control AGV system. 
\end{abstract}


\section{Introduction} 
New use-cases have emerged in vertical industries with the launch of 5G technology. Applications such as logistics, assembly lines and automotive manufacturing have growing demands for cloud control AGVs. Traditional AGVs are programmed to execute a designated task autonomously. As the controllers are located on the AGV, path-planning and task-customization is painstaking. Therefore to facilitate more flexibility and agility, a cloud based controller is proposed for new generation of AGVs. The cloud control AGV enables on-demand task updating and path planning while simultaneously collaborating and coordinating with the other connected devices in a factory. The energy consumption of an AGV can also be reduced by selectively offloading the controller function in cloud \cite{tayade2017device,tayadeICC}. Extensive research on the coordinated AGV controller is performed in recent years \cite{aguirre2011remote,Kanjanawanishkul,multiplerobots,olmi2008coordination}. The communication technologies are enhanced to enable cloud based industrial use-cases in \cite{ pub11081, 8539695,9162929}. The paper \cite{Kanayama1990}, presents a continuous-time stable AGV control system with limited velocity and acceleration, however, in practice the AGV controller operates in discrete time. Therefore, to study the error performance of a cloud based AGV system, we discretize the continuous-time control system presented in \cite{Kanayama1990}. A key characteristic of a controller is to assure stability of a system. The stability of a system is evaluated from the system response. A system is considered stable if the response of system is either bounded or decreases to zero over time. The AGV control system is time-varying and non-linear, hence, the generic approach to evaluate the stability criterion of a LTI system is inapplicable \cite{nof2009springer,inbook}.
Moreover, in a cloud control AGV system, the communication between a controller and an actuator occurs over an imperfect wireless channel. The wireless link may introduce a delay or loss of the information packets causing instability of the control system. The research in \cite{1310480NoisyChannel,1333206stochasticcontrol,  AGVSCC, CodingRateShreya} investigates the impact of wireless network uncertainties on the control system. The authors in \cite{1310480NoisyChannel, 1333206stochasticcontrol} considered a basic LTI system, whereas many practical control systems are non-linear time-varying in nature.  \cite{AGVSCC, CodingRateShreya} presents the impact of downlink channel outages on the stability of a non-linear time-varying system, it does not consider uplink channel outages. In this paper we evaluate the stability criterion of a non-linear, discrete, time-varying AGV system in presence of uplink channel outages. Furthermore, the authors in \cite{Kanayama1990} determine the control rule and optimal feedback parameters to maintain AGV's stability. In this paper, we analyse the impact of feedback control parameters on the error convergence of a cloud based AGV system. The error convergence of an AGV is analysed for two different scenarios; with no limitation on the AGV velocity and with maximum permissible AGV velocity.

The objective of this paper is to evaluate the stability of a discrete non-linear, time-varying, in-homogeneous AGV control system taking into account uplink channel outages. An approach presented in \cite{nof2009springer} is followed to evaluate the stability of a non-linear time varying control system. In Section~\ref{sec:Ap0:1}, we describe the discretization process of a continuous-time AGV control system. The control rule and stability in presence of uplink channel outages is evaluated in Section~\ref{sec:Ap0:10} and  Section~\ref{sec:Ap0:10} respectively. The results on error convergence rate with respect to feedback parameters are presented in Section~\ref{sec:Ap0:15}. The conclusions are discussed in Section~\ref{sec:conclusion}.



\begin{figure*}[h!] 
\centering
\includegraphics[width =0.9\textwidth]{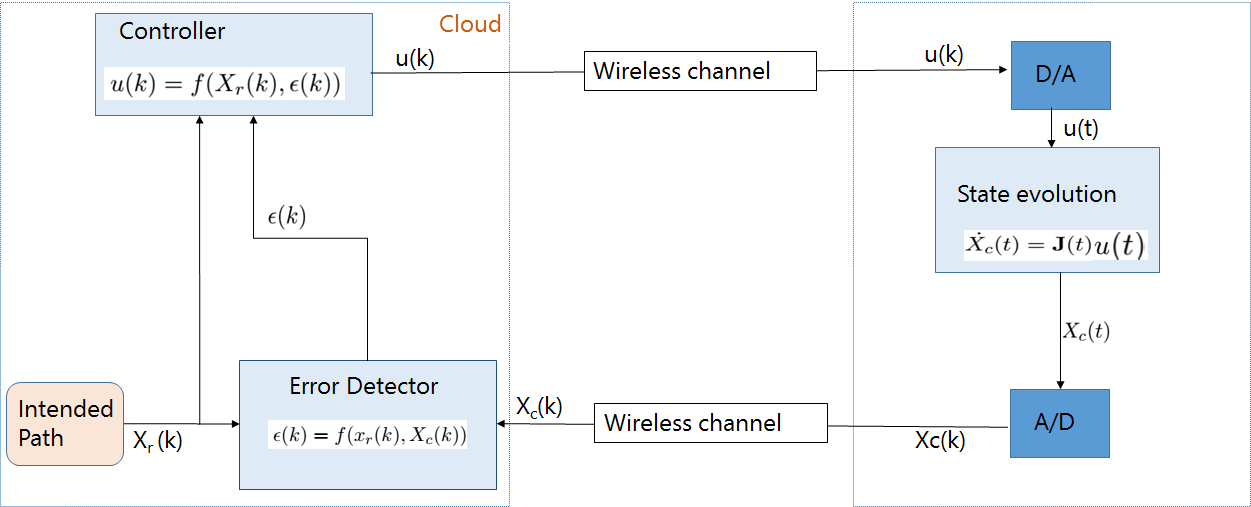}
\caption{Discrete system of Cloud Controlled AGV}
\label{fig:ch3:1}
\end{figure*}

\section{AGV control system} \label{sec:Ap0:1}
An AGV follows an intended reference path $X_r =[x_r, y_r, \theta_r]$ where $(x_r,y_r)$ is the x-y coordinate and $\theta$ is the angle of orientation. The Fig.~\ref{fig:ch3:1}, shows the basic function blocks of the control system. The error detector evaluates the difference between the actual AGV position and the intended path. The controller generates translational $\nu$ and rotational velocity $\omega$ according to the evaluated position error. These velocities are the control inputs applied to an AGV actuator. The AGV attains a new position $X_c$ = [$x_c$, $y_c$, $\theta_c$] following the application of control inputs. The updated AGV position is feedback to the error detector. The error is sent to the controller and new control inputs are determined. The process continues until the reference path is completed by the AGV. The feedback mechanism direct the error to approach zero, as time goes to infinity. 

The paper \cite{Kanayama1990} present a continuous-time AGV control system. However, for a cloud controlled AGV, the control inputs and the AGV's position updates are received at discrete time instants over a wireless channel. In this section, a detailed functional description and the mathematical model for each block of a discrete AGV control ~\ref{fig:ch3:1} is presented. 

\paragraph{Error detector} The error detector evaluates the error $\epsilon(k)$ between the actual position $X_c$ of an AGV, to the reference position $X_r$ at $k^{th}$ time sample as
\begin{align}
    \epsilon(k) &= \left(\begin{array}{c}x_e(k)\\y_e(k)\\\theta_e(k)\end{array}\right)  \\
    &= \left(\begin{array}{ccc}\cos\theta_c(k) & \sin\theta_c(k) & 0\\-\sin\theta_c(k) & \cos\theta_c(k) & 0\\0 & 0 & 1\end{array}\right) \left(X_r(k) - X_c(k)\right)  \\
     &= \mathbf{T_e}(k) \left(X_r(k) - X_c(k)\right), \label{eq:ch3.ieee.10}
\end{align}
where $\mathbf{T_e(k)}$ is the rotational matrix, $x_e$, $y_e$ is the error determined as a result of difference in x, y coordinate, and $\theta_e$ is the error in the orientation of an AGV.

\paragraph{Controller} The controller calculates the control input $u(k)$ = [$\nu(k); \omega(k)$] according to the error $\epsilon(k)$ determined in \eqref{eq:ch3.ieee.10} as
\begin{align}
    u(k) &= \left[\begin{array}{c}\nu(k)\\ \omega(k)\end{array}\right] \nonumber \\
    &= \left[\begin{array}{c} \nu_r(k) \cos\theta_e(k) + K_x x_e(k)\\ \omega_r(k) + \nu_r(k) \left[ K_y y_e(k) + K_\theta \sin\theta_e(k) \right] \end{array}\right], \label{eq:ch3.ieee.15}
\end{align}
where $K_x\,[\unit{s^{-1}}]$, $K_y\,[\unit{m^{-1}}]$ and $K_\theta\,[\unit{m^{-1}}]$ are constants which impact the convergence rate of the control system. The stabilized control rule without uplink channel outages is derived in \cite{Kanayama1990}, in the section~\ref{sec:Ap0:6}, we evaluate a stable control rule with uplink channel outages. 

\paragraph{State evolution of an AGV} Following the application of the control inputs $\nu$ and $\omega$, an AGV attains a new position state. The position state evolution of an AGV is described in differential form as
\begin{align}
\dot{X_c}(t) &=\left(\begin{array}{c}\dot x_c(t)\\ \dot y_c(t)\\ \dot \theta_c(t)\end{array}\right)  = \underbrace{\left(\begin{array}{cc} \cos\theta_c(t)& 0\\ \sin\theta_c(t) & 0 \\ 0 & 1\end{array}\right)}_{\mathbf{J}(t)} u(t) \\
\dot{X_c}(t) &= \mathbf{J}(t) u(t), \label{eq:AGV:320}
\end{align}
 where $\dot{x}_c$ = $\nu_x$ = $\cos\theta_c(t) \cdot \nu(t)$ is the velocity component in x-direction. The actual position of an AGV is obtained by solving ~\eqref{eq:AGV:320}. 
 
As shown in the Fig.~\ref{fig:ch3:1}, controller at the cloud generates control input $u(k)$ at time $t_k = k \cdot T_s$ where $k = \left\lbrace 0, 1 \dots \right\rbrace$, represent the sample point and $T_s$ as the sampling time. These control inputs $u(k)$ are sent to the actuator of an AGV over a wireless channel. The discrete control inputs cannot be directly applied to the actuator of an AGV, as the AGV state evolution is a physical process and continuous in time ~\eqref{eq:AGV:320}. The state of an AGV will continuously evolve for the given control input. Therefore, the control inputs are converted into the continuous time by D/A converter. The D/A converter holds the control input $u(k)$ for the sampling time period $T_s$, before it is applied to an actuator. The position state $X_c$ of an AGV's actuator evolves continuously for the given control inputs $u(k)$, although the control inputs $u(k)$ are constant over a sampling time period $T_s$. Furthermore, the current position updates of an AGV  are feedback to the controller. These discrete updates are obtained by sampling the position state at $ k^\text{th}$ time instant by A/D converter.  
 
\section{Discretization} \label{sec:Ap0:5} As the exchange of information between the AGV actuator and cloud controller is in discrete time, it is necessary to evaluate the behavior of the state variables $X_c(t)$ at these discontinuities. The actual position state $X_c$ of an AGV at time $t$, is obtained by solving the differential equation in ~\eqref{eq:AGV:320}. The control inputs are constant sampling time period $T_s$ = $t_\text{k+1}$-$t_\text{k}$, the state at any time $t$, such as $t_k < t < t_{k+1}$  can be evaluated by integrating the above equation over the time interval $\left[t_{k} \: t \right]$.

\begin{equation}
\int\limits^{t}_{t_{k}} \dot{X}_c(t) dt = \int\limits^{t}_{t_{k}} J(t) u(t) dt 
\end{equation}
$u(t)$ is constant over the given sampling time period, hence, $u(t)$ =  $u(t_k)$, is independent of dt. 
\begin{equation}
X_c(t) =  X_c(t_k) + \int\limits^{t}_{t_{k}} J(t) dt \cdot  u(t_k)
\end{equation}
where $J(t)$ is a in-homogeneous non-linear system matrix, hence, the integration cannot be solved in a closed form. Therefore, to evaluate the approximate solution to the differential equation, we use piece-wise integration approach.
Assuming that the sampling interval $T_s = t_{k+1} - t_k$, is small, $J(t)$ will be linear over this time interval. The equation transforms to 
\begin{eqnarray}
X_c(t) =  X_c(t_k) +  \int\limits^{t}_{t_{k}} J(t) dt \cdot  u(t_k) \\
X_c(t) =  X_c(t_k) +  (t-t_k) \cdot J(t_k) \cdot  u(t_k)
\end{eqnarray}
The position state variable $X_c(t)$ for a sufficiently low sampling period, at time  $t = t_{k+1}$ is determined as
\begin{eqnarray}
X_c(t_{k+1}) =  X_c(t_k) +  T_s \cdot J(t_k) \cdot  u(t_k)
\end{eqnarray}
The time $t_k$ can be represented in terms of samples as 
\begin{equation}
X_c(k+1) =  X_c(k) +  T_s \cdot J(k) \cdot  u(k)
\end{equation}

\begin{align}
\underbrace{\left[\begin{array}{c} x_c(k+1)\\  y_c(k+1)\\ \theta_c(k+1)\end{array}\right]}_{X_c(k+1)} &= \underbrace{\left[\begin{array}{c} x_c(k)\\  y_c(k)\\ \theta_c(k)\end{array}\right]}_{X_c(k)} + T_s \cdot \mathbf{J}(k) \underbrace{\left[ \begin{array}{c} \nu(k) \\ \omega(k)\end{array} \right]}_{u(k)}
\end{align}
where $u(k)$ is the control input at $k-\text{th}$ sample, and $\mathbf{J}(k)$ is 
\begin{align}\label{eq:ch3.ieee.19}
\mathbf{J}(k) = \left(\begin{array}{cc} \cos\theta_c(k)& 0\\ \sin\theta_c(k) & 0 \\ 0 & 1\end{array}\right). 
\end{align}
Here, $\theta_c(k)$ is the orientation of an AGV with respect to the x-axis at the $k^\text{th}$ sample. Therefore, the equation becomes
\begin{equation}
X_c(k+1) =  X_c(k) +  T_s \cdot \mathbf{J}(k) \cdot u(k). \label{eq:ch3.ieee.20} 
\end{equation}
The solution in ~\eqref{eq:ch3.ieee.20} is the discrete approximate to evaluate the position state of an AGV at next time instant. The method followed is the Euler's method to solve nonlinear differential equations. The evaluated position state of an AGV is sent to the error detector to evaluate the error.

\section{Impact of uplink channel outages} \label{sec:Ap0:6}
\paragraph{Control rule}
The control rule in ~\eqref{eq:ch3.ieee.15} does not consider the impact of wireless channel outages. In this section a control rule and stability criterion are determined with uplink channel outages. For the sake of brevity, downlink channel is assumed perfect. The position update $X_c(k)$ is transmitted over uplink channel at every $T_s$ seconds. In presence of an uplink channel outages, the position update $X_c(k)$ is not successfully received by an error detector at the cloud. The error is therefore calculated with the outdated AGV position $X_c(k-n_{ul})$, where $n_{ul}$ is the number of consecutive channel outages. The error with the outdated  position update $X_c(k-n_{ul})$ is 
\begin{align}\label{eq:AGV:1000}
 \epsilon_{ul} &=  \mathbf{T_e}(k-n_{ul}) \left(X_r(k) - X_c(k-n_{ul}) \right) \\
 \left[\begin{array}{c}x_{e,ul}(k)\\y_{e,ul}(k)\\\theta_{e,ul}(k)\end{array}\right] &= \left(\begin{array}{ccc}\cos\theta_c(k-n_{ul}) & \sin\theta_c(k-n_{ul}) & 0\\-\sin\theta_c(k-n_{ul}) & \cos\theta_c(k-n_{ul}) & 0\\0 & 0 & 1\end{array}\right) \times \nonumber \\
    &  \left[X_r(k) - X_c(k-n_{ul})\right]
\end{align}
where $\mathbf{T_e(k)}$ is the rotational matrix and $\epsilon_{ul}$ is the error with uplink channel outages. The new control rule $u_{ul}$ is generated by substituting the error from ~\eqref{eq:AGV:1000} in ~\eqref{eq:ch3.ieee.15} as 
\begin{align}
    u_{ul}(k) &= \left[\begin{array}{c}\nu_{ul}(k)\\ \omega_{ul}(k)\end{array}\right] \nonumber \\
    &= \left[\begin{array}{c} \nu_r(k) \cos\theta_{e,ul}(k) + K_x x_{e,ul}(k)\\ \omega_r(k) + \nu_r(k) \left[ K_y y_{e,ul}(k) + K_\theta \sin\theta_{e,ul}(k) \right] \end{array}\right]
\end{align}
These control inputs $u_{ul}(k)$ are applied to an AGV actuator therefore new position state at time $t_{k+1}$ is 

\begin{equation} \label{eq:Ap0:100}
 X_c(k+1) =  X_c(k) +  T_s \cdot \mathbf{J}(k) u_{ul}(k)
\end{equation}

\begin{figure*}[t!]
\begin{align}\label{eq:AGV:365}
  \dot{\epsilon}(t) = 
  \left(\begin{array}{c}
	(\omega_r(t) + \nu_r(t) \left[ K_y y_e(t) + K_\theta \sin\theta_e(t) \right])y_e(t) - \nu(t) + \nu_r(t)\cos\theta_e(t)\\
	-(\omega_r(t) + \nu_r(t) \left[ K_y y_e(t) + K_\theta \sin\theta_e(t) \right]) x_e(t) + \nu_r(t)\sin\theta_e(t)\\ \omega_r(t) - \omega_r(t) + \nu_r(t) \left[ K_y y_e(t) + K_\theta \sin\theta_e(t) \right]
\end{array}\right) 
\end{align}
\end{figure*} 
\begin{figure*}
\begin{equation}\label{eq:ch3.ieee.55}
\resizebox{.9\hsize}{!}{$\mathbf{A}(k) = \left(\begin{array}{ccc} 1- T_s \cdot K_x \cdot \cos \left[\theta_c(k) \right] \cdot \cos\left[\theta_c(k-n_{ul})\right] & - T_s\cdot K_x \cdot \cos\left[\theta_c(k)\right] \cdot \sin\left[\theta_c(k-n_{ul}) \right] & - T_s \sin\left[\theta_c(k) \right] \nu_r(k-n_{ul}) \\ 
-T_s \cdot K_x \cdot \sin\left[\theta_c(k)\right] \cdot \cos\left[\theta_c(k-n_{ul}) \right] & 1-T_s\cdot K_x \cdot \sin \left[\theta_c(k)\right] \cdot \sin\left[\theta_c(k-n_{ul})\right] & T_s \cdot \cos \left[ \theta_c(k)\right] \nu_r (k-n_{ul}) \\ T_s \cdot k_y \sin\left[\theta_c(k-n_{ul})\right] \nu_r(k-n_{ul}) & - T_s \cdot k_y \cos\left[\theta_c(k-n_{ul})\right] \nu(k-n_{ul}) & 1-T_s\cdot k_\theta \nu_r(k-n_{ul})\end{array}\right)$}
\end{equation}
\end{figure*}

\paragraph{Error dynamics and Stability} \label{sec:Ap0:10}
The control rule in ~\eqref{eq:ch3.ieee.15} is derived such that it can maintain system stability in presence of an error. A system is stable around $\epsilon=0$, if the error $\epsilon$ $\to$ 0 or is bounded. Therefore, we analyse the error convergence by evaluating the error dynamics of an AGV control. The error dynamics of the continuous time system is evaluated by taking the time derivative of error as 
\begin{eqnarray}\label{eq:AGV:400}
    \dot{\epsilon}(t) = \left(\begin{array}{c}
	\omega(t)y_e(t) - \nu(t) + \nu_r(t)\cos\theta_e(t)\\
	-\omega(t)x_e(t) + \nu_r(t)\sin\theta_e(t)\\
	\omega_r(t) - \omega(t) 
      \end{array}\right),
  \end{eqnarray} \cite{Kanayama1990}.
Substituting the control inputs $\omega$ and $\nu$ from \eqref{eq:ch3.ieee.15},  \eqref{eq:AGV:365} is derived. The non-linear error dynamics in \eqref{eq:AGV:365} are linearized to the form of ~\eqref{eq:AGV:440}, by taking the Jacobian of differential equation at $\epsilon$ = 0, 
\begin{align} \label{eq:AGV:440}
\dot{\epsilon}(t) = A(t) \epsilon(t)
\end{align} 
The system is stable if all the real roots of characteristics equation for $A$ are negative as per the Routh-Hurwitz Criterion. The stability is already proved in \cite{Kanayama1990}. 

\paragraph{Stability with uplink outages}
The error dynamics of a discrete time control system should be known to evaluate the stability criterion with uplink channel outages. The position state evolution with uplink channel outages is derived in ~\eqref{eq:Ap0:100}. An equilibrium is reach if the error $\epsilon_{ul}(k) = 0$, i.e\, $X_r(k+1)$ - $X_c(k+1)$ = 0. We linearize the state space system of ~\eqref{eq:Ap0:100} in the form $X_c(k+1) = \mathbf{A}(k) X_c(k) + \mathbf{B}(k) u_{ul}(k)$ by taking Jacobian of \eqref{eq:Ap0:100} and then substituting $X_r = X_c$. The control matrix $\mathbf{A}$ is evaluated in ~\eqref{eq:ch3.ieee.55}, by taking partial derivatives of \eqref{eq:Ap0:100} with respect to $x_c$, $y_c$ and $\theta_c$ respectively \cite{AGVSCC}. If $\{\lambda_1(k), \dots, \lambda_M(k)\}$ = Eig($\mathbf{A(k)}$) are the  Eigenvalues of $\mathbf{A}$, a (LTI) system is stable if
\begin{equation}\label{eq:ch3.ieee.56}
    \forall i,k: 0 < |\lambda_i(k)| < 1
\end{equation}  \cite{Astrom:1990:CST:78995, nof2009springer}. To check the stability of a time varying system, the test for stability needs to be repeated for each time step $k$ as evaluated for downlink channel outages in \cite{AGVSCC, CodingRateShreya}.




\begin{figure*}[h!]
\begin{subfigure}{0.5\textwidth}
\centering
\includegraphics[width=0.7\linewidth]{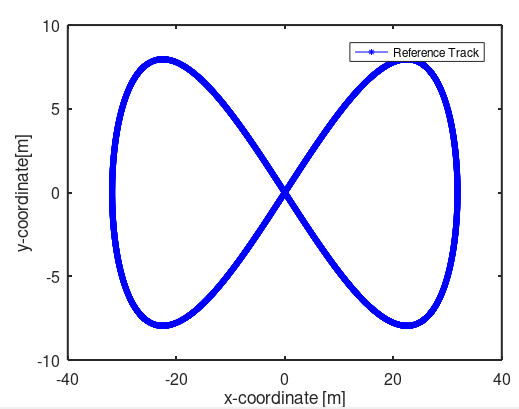}
\caption{AGV reference track $X_r$} 
\label{fig:Ap0.Result.1}
\end{subfigure}
\hfill
\begin{subfigure}{0.5\textwidth}
\includegraphics[width=0.7\textwidth]{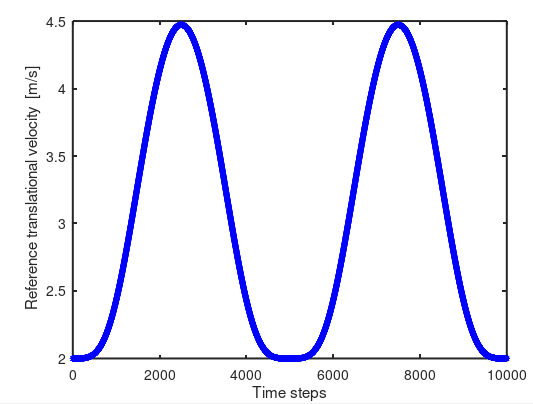}
\caption{Reference translational velocity $\nu_r$} 
\label{fig:Ap0.Result.2}
\end{subfigure}
\caption{Reference track and velocity}
\end{figure*} 

\section{Simulation setup and Results}\label{sec:Ap0:15}
\subsection{Simulation setup}
An AGV traces a reference track $X_r$ shown in Fig.~\ref{fig:Ap0.Result.1}, in $T$ = \unit[100]{s}. The number of steps required to trace the complete track is $N_{timesteps}$ = $T/T_s$, where $T_s = \unit[5]{ms}$. The reference track and the reference velocity $\nu_r$ for each time step is shown in Fig.~\ref{fig:Ap0.Result.2}. In presence of non-zero error, the control inputs are generated according to ~\eqref{eq:ch3.ieee.15} and are applied to the AGV actuators so that the error $\epsilon(k)$ $\to$ 0, as $k \to N_{timesteps}$. The optimal selection of feedback parameters $K_x$,$K_y$ and $K_\theta$ is required to immediately reduce the error. The discrete AGV controller is implemented in Matlab.

\begin{figure*}[t!]
\begin{subfigure}{0.3\textwidth}
\centering
\includegraphics[width=1\textwidth]{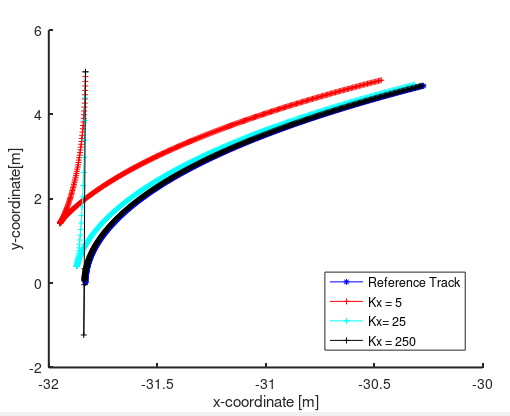}
\caption{AGV's followed tracks: unlimited $\nu$}
\label{fig:Ap0.Result.3}
\end{subfigure}
\begin{subfigure}{0.3\textwidth}
\centering
\includegraphics[width=1\textwidth]{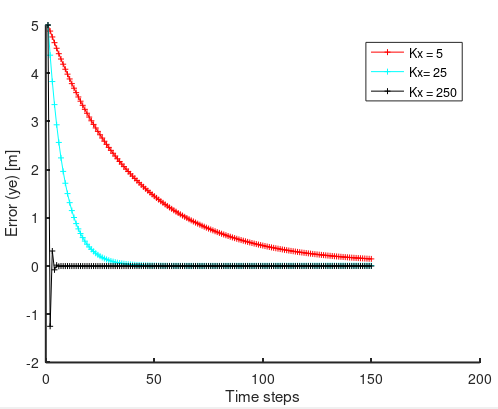}
\caption{Error in y-direction $y_e$: unlimited $\nu$} 
\label{fig:Ap0.Result.4}
\end{subfigure}
\begin{subfigure}{0.3\textwidth}
\includegraphics[width=1\textwidth]{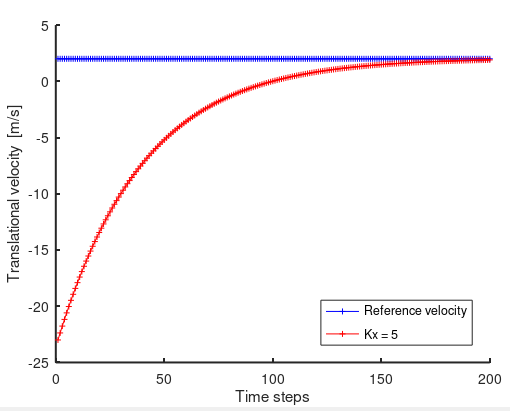}
\caption{Translational velocity: unlimited $\nu$} 
\label{fig:Ap0.Result.5}
\end{subfigure}
\begin{subfigure}{0.3\textwidth}
\includegraphics[width=1\textwidth]{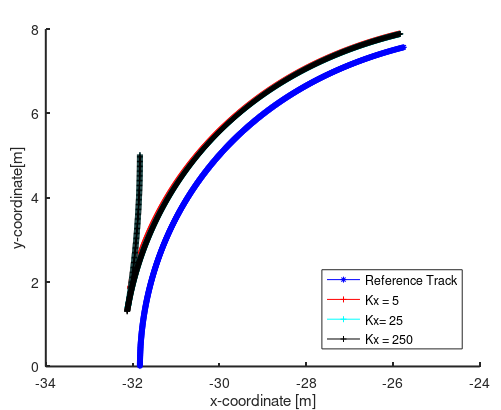}
\caption{AGV's followed tracks: limited $\nu$} 
\label{fig:Ap0.Result.6}
\end{subfigure}
\begin{subfigure}{0.3\textwidth}
\includegraphics[width=1\textwidth]{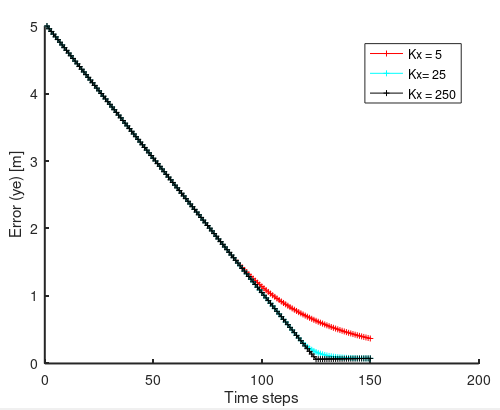}
\caption{Error in y-direction $y_e$: limited $\nu$} 
\label{fig:Ap0.Result.7}
\end{subfigure}
\begin{subfigure}{0.3\textwidth}
\includegraphics[width=1\textwidth]{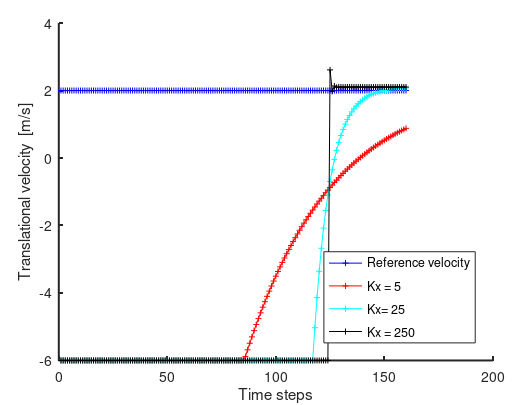}
\caption{Translational velocities: limited $\nu$} 
\label{fig:Ap0.Result.8}
\end{subfigure}
\caption{AGV tracks and error convergence}
\end{figure*}

\begin{table}[h!]
 \caption{Variable description}
  \label{tb:notation}
  \centering
  \begin{tabular}{|c||c|}
    \hline 
    Variable & Description \\ \hline \hline
    $X_r$ & Reference position vector of an AGV  \\ \hline 
      $X_c$ & Actual position vector of an AGV \\ \hline
       $\nu$ & Translational velocity \\ \hline
        $\omega$ & Rotational velocity \\ \hline
        $\nu_r$ & Reference translational velocity \\ \hline
        $\omega_r$ & Reference rotational velocity \\ \hline
        $u$ & Control information vector \\ \hline
        $T_s$ & Sampling time \\ \hline
        $T$ & time to trace complete reference track \\\hline
        $N_{timesteps}$ & Total number of timesteps \\ \hline
        $k$ & Time step\\ \hline
        $\epsilon$ & Position error\\ \hline
        $x_e$ & Position error in x-direction\\ \hline
        $y_e$ & Position error in y-direction\\ \hline
        $\theta_e$ & Orientation error\\ \hline
        $n_{ul}$ & Consecutive uplink channel outages\\ \hline $\epsilon_{ul}$ &  Error with uplink channel outages   \\ \hline
         $u_{ul}$ &  Control inputs with uplink channel outages   \\ \hline
\end{tabular}
\end{table}

\subsection{Impact of feedback control parameters on error convergence}
To study the impact of feedback control parameters on the error convergence, an error in y-direction, $y_e$ = \unit[5]{m} is introduced. The controller generates $\nu$ and $\omega$ according to ~\eqref{eq:ch3.ieee.15}. The parameter $1/K_x$ is the time constant that directs the rate at which error decreases to 0. In the discrete control, the parameter $K_x$ determines the number of steps required to direct an AGV back on the reference track. 

\paragraph{No limitation on the translational velocity $\nu$ }
The Fig.~\ref{fig:Ap0.Result.3}, shows the starting position of an AGV is $\unit[5]{m}$ away from the reference position in  y-direction at time step $k=1$. The Fig.~\ref{fig:Ap0.Result.3}, shows the tracks followed by AGV when $K_x$ is increased from $\unit[5]{/ sec}$ to $\unit[250]{/sec}$. At lower $K_x$ more number of steps are required for an AGV to return to the reference track. As $K_x$ is increased to $\unit[25]{/ sec}$, the convergence rate increases. On further increasing the $K_x = \unit[250]{/sec}$, an AGV oscillate before returning to the reference track. The optimal response of an AGV is with no oscillation and higher rate of error convergence. The rate of convergence of y-directional error for varying $K_x$ is shown in Fig.\ref{fig:Ap0.Result.4}. Furthermore,  Fig.~\ref{fig:Ap0.Result.5}
shows the required velocity at every time step to reduce the error at the rate specified by $K_x$. The required $\nu = \unit[-25]{m/s}$, from time-step $k= 1$ to $k = 20$, with $K_x = \unit[5]{/sec}$ is very high for an AGV operating on shop floor. Application of such high velocities is not practically feasible, although the error converges in less number of timesteps. Therefore, we evaluate the behaviour of error convergence with limited $\nu$. 

\paragraph{Limitation on the translational velocity $\nu \leq \unit[6]{m/s}$ }
The Fig.~\ref{fig:Ap0.Result.6}, Fig.~\ref{fig:Ap0.Result.7} and Fig.~\ref{fig:Ap0.Result.8}, shows the AGV tracks, error convergence and the translational velocity required to navigate an AGV back on the reference track by driving the AGV with maximum permissible velocity of $\unit[6]{m/s}$. Due to limitation on the AGV velocity, feedback parameter $K_x$ does not have a substantial impact on the error convergence. The error convergence with $K_x = \unit[25]{/sec}$ and $K_x = \unit[250]{/sec}$, shows similar behaviour of error over time. The response of translational velocity of an AGV is shown in Fig.~\ref{fig:Ap0.Result.8}. At lower $K_x$ the generated control inputs, i.e.\, the $\nu$, gradually approaches the reference velocity. It illustrates a case of over-damped response,where the AGV returns to the reference track gradually and requires more number of time-steps. On contrary, at higher $K_x = \unit[250]{/sec}$, $\nu$ shows under-damped response, it converges by oscillating around the reference velocity, which is not the optimal case.  If the oscillations are higher, it takes more time to converge. An optimal case is of critical damping i.e.\, $\nu$ converges with less number of timesteps and without oscillation. The optimal convergence is realized at $K_x = \unit[25]{/sec}$.

\section{Conclusions}\label{sec:conclusion}
We consider a practical use-case of an edge cloud AGV control system. A discretization process of a stable continuous time AGV control system is presented. We evaluate the control rule, error dynamics and stability criterion of a discrete cloud control AGV in presence of uplink channel outages. The derived stability criterion shows that the stability of an AGV control is dependent upon the number of consecutive uplink outages and the Eigen values of control matrix $A$ \eqref{eq:ch3.ieee.55}. Furthermore, we evaluate optimal feedback control parameters that increases the rate of error convergence. The paper consider two scenarios to study the impact of feedback parameters on error convergence; with no limitation on AGV velocity and with limited AGV velocity to \unit[6]{m/s}. The result reveals that the optimal feedback parameter is chosen with critical damping, i.e.\, it shows no oscillation an higher rate of convergence. For cloud control AGV the critical damping is achieved at $K_x = \unit[25]{/sec}$. Moreover, the results reveal that the impact of feedback control parameters is higher in case of no-limitation on the AGV velocity. If the AGV velocity is limited, the rate of error convergence does not vary significantly with varying $K_x$. This effect is seen due to limitation on the control inputs i.e.\, AGV velocity applied to an AGV and the optimal feedback parameters is chosen from the response of the control inputs. The feedback parameter corresponding to fastest convergence of control inputs to the reference velocity with no oscillation is the optimal parameter. 
The paper investigate the impact of feedback parameter corresponding to an error in y-direction. In future, the research can be extended to evaluate the optimal feedback parameters for x-direction and the orientation of an AGV. The impact of uplink outages on the error convergence can also be investigated.

\bibliographystyle{IEEEtran} 
\bibliography{IEEEabrv, SCC.bib, bibilography.bib}
\end{document}